\documentclass{nature}
\usepackage{graphicx}
\usepackage{amsmath}
\usepackage{amssymb}
\usepackage{textcomp}
\usepackage{gensymb}
\usepackage{xcolor}
\usepackage{lineno}
\usepackage{soul}
\usepackage{colortbl}
\usepackage{ulem}
\usepackage[labelfont=bf]{caption}
\usepackage{multibib}
\newcites{supp}{Supplementary References}
\makeatletter
\let\saved@includegraphics\includegraphics
\AtBeginDocument{\let\includegraphics\saved@includegraphics}
\renewenvironment*{figure}{\@float{figure}}{\end@float}
\makeatother
\newcommand{\reftextit}[1]{}
\usepackage{graphicx}

\title{Field-Tunable Valley Coupling and Localization in a Dodecagonal Semiconductor Quasicrystal}
\author{~Zhida Liu$^{1,2}$,~Qiang Gao$^{1,2,3}$, ~Yanxing Li$^{1,2}$,~Xiaohui Liu$^{1,2}$,~Fan Zhang$^{1,2}$,
~Dong Seob Kim$^{1,2}$,~Yue Ni$^{1}$,~Miles Mackenzie$^{1}$,~Hamza Abudayyeh$^{1}$,
~Kenji Watanabe$^{4}$,~Takashi Taniguchi$^{5}$,~Chih-Kang Shih$^{1,2}$, 
~Eslam Khalaf$^{1, 3*}$, and~Xiaoqin Li$^{1,2*}$}

\begin{document}
\maketitle
{\renewcommand{\baselinestretch}{1.5}
\begin{affiliations}
\item Department of Physics and Center for Complex Quantum Systems, The University of Texas at Austin, Austin, Texas, 78712, USA.
\item Center for Dynamics and Control of Materials and Texas Materials Institute, The University of Texas at Austin, Austin, Texas, 78712, USA.
\item Department of Physics, Harvard University, Cambridge, Massachusetts 02138, USA.
\item Research Center for Electronic and Optical Materials, National Institute for Materials Science, 1-1 Namiki, Tsukuba 305-0044, Japan.
    \item Research Center for Materials Nanoarchitectonics, National Institute for Materials Science, 1-1 Namiki, Tsukuba 305-0044, Japan.

 \thanks{
  Corresponding author:~\textcolor{blue}{{elaineli@physics.utexas.edu}, {eslam\_khalaf@fas.harvard.edu}}}
\end{affiliations}
}
{\renewcommand{\baselinestretch}{1.7}

\date{\today}
\pagebreak

\begin{abstract}{
Quasicrystals are characterized by atomic arrangements possessing long-range order without periodicity. Van der Waals (vdW) bilayers provide a unique opportunity to controllably vary atomic alignment between two layers from a periodic moir\'e crystal to an aperiodic quasicrystal. Here, we reveal a remarkable consequence of the unique atomic arrangement in a dodecagonal WSe$_2$ quasicrystal: the $K$ and $Q$ valleys in separate layers are brought arbitrarily close in momentum space via higher-order Umklapp scatterings. A modest perpendicular electric field is sufficient to induce strong interlayer $K-Q$ hybridization, manifested as a new hybrid excitonic doublet. Concurrently, we observe the disappearance of the trion resonance and attribute it to quasicrystal potential driven localization. Our findings highlight the remarkable attribute of incommensurate systems to bring any pair of momenta into close proximity, thereby introducing a novel aspect to valley engineering.}

\end{abstract}
}

\maketitle
\renewcommand{\baselinestretch}{2}
\setlength{\parskip}{7pt}
Quasicrystals, first discovered in metallic alloys four decades ago, have redefined the concept of crystals~\cite{shechtman1984metallic,levine1986quasicrystals,levine1984quasicrystals}. One prominent distinction between quasicrystals and their crystalline counterparts is that the absence of translational symmetry precludes the existence of Bloch waves in quasicrystals. As a result, quasi-periodic potentials can mimic the properties of random disorder leading to single-particle (and potentially many-body) localization physics~\cite{biddle2009localization,an2021interactions,mastropietro2015localization} as well as other unusual transport, thermal, and magnetic properties\cite{dolinvsek2012electrical,gautier2021strongly,else2021quantum}. However, beyond photonic and cold atom systems~\cite{man2005experimental,matsui2007transmission,gellermann1994localization,jeon2017intrinsic,wang2024observation,shimasaki2024anomalous}, intrinsic properties of quasicrystals have been very difficult to study in solid state systems because the inevitable disorder in the chemical composition and structure of realistic materials often obscure the pertinent observations.

Twisted vdW bilayers and multilayers provide a completely new platform to study quasicrystals that overcome these hurdles. In vdW bilayers, atomic arrangements are known and tunable between aperiodic (i.e. quasicrystal) and periodic (moir\'e crystal) readily by controlling the twist angle without introducing additional disorder~\cite{zhao2021strong,carr2020electronic,koren2016coherent,zhang2020twist}. A few experiments have been performed on graphene dodecagonal quasicrystal bilayers, where Dirac cone replica with the 12-fold rotational symmetry were observed in angle-resolved photoemission spectra~\cite{ahn2018dirac,yao2018quasicrystalline}. However, due to the large twist angle, combined with the linear dispersion of graphene and its large Fermi velocity, significant interlayer coupling only takes place at high energies leaving essentially decoupled Dirac cones at the Fermi level~\cite{pezzini202030,yu2019dodecagonal,deng2020interlayer,moon2019quasicrystalline}. The absence of electronic coupling near the Fermi level has thus limited the influence of these unique single-particle electronic states on the transport and electronic properties of graphene bilayers. In contrast, the electronic bands from transition metal dichalcogenides (TMDs) feature a significantly shallower dispersion enabling interlayer coupling at large twist angles to be significant at energies close to the Fermi surface. This is manifested through the appearance of a set of dense mini-gaps near the valence band maximum as recently revealed by scanning tunneling microscopy~\cite{yanxing2024Tuning}. While these prior studies have established the possibility of creating artificial tunable quasicrystals based on twisted vdW bilayers, most special implications enabled by this novel quasicrystal platform remain unexplored.
  
Here, we unveil a surprisingly strong coupling between two distinct valleys ($K$ and $Q$) in a dodecagonal quasicrystal formed in a 30$^\circ$ WSe$_2$ bilayer, enabled by the incommensurate atomic configurations. 
While interlayer hybridization is common in stacked vdW layers, it typically occurs between states with substantial overlap in crystal momentum and energy~\cite{barre2024engineering}. In quasicrystals, the unique coupling between $K$ and $Q$ valleys in separate layers is facilitated by the exceptionally efficient higher-order Umklapp scatterings in quasicrystals, and manifested as the emergence of new exciton resonances when a modest E-field further induces interlayer hybridization. Concurrently, the trion resonance disappears, which we attribute to quasicrystal-induced localization, a long-sought-after hallmark of quasicrystals. These spectroscopic features are completely absent in a 21.8$^\circ$ moir\'e crystal that retains translational symmetry. Our studies suggest that vdW bilayers offer a controllable platform for exploring intrinsic quasicrystal properties inaccessible in other systems including commensurate moir\'e superlattices.

Generally speaking, a key qualitative difference between the commensurate and incommensurate systems is that, for the latter, one can always find an Umklapp process of sufficiently higher order to bring any pair of momenta arbitrarily close. In a 30$^\circ$ WSe$_2$ bilayer, the relevant scattering connects the $K$ valley of the top layer and $Q$ valley of the bottom layer, which can be further energetically aligned by applying a modest upward E-field. The atomic arrangements of the WSe$_2$ dodecagonal quasicrystal and 21.8$^\circ$ twisted moir\'e superlattice are illustrated in Fig.~1a where the red (grey) dots represent W (Se) atoms, respectively. In the latter case, the moir\'e superlattice features periodic supercells with a lateral size $\sqrt7~a_0$, where $a_0$ represents the lattice constant of WSe$_2$ monolayer. In contrast, the quasicrystal breaks the translation symmetry but retains a 12-fold rotational symmetry.  
We examine the reciprocal space of the quasicrystal closely in Fig.~1b, where the red (blue) hexagonal pattern represents the extended Brillouin Zone (BZ) of the top (bottom) layer, respectively. 
In the 30$^\circ$ quasicrystal, $Q$ valleys in the bottom layer roughly lie at $|\Gamma-Q_b| =0.60~|\Gamma-K_b|$ ~\cite{debnath2022tuning}. We label the top layer $K$ valleys ($K_t$) in the first few BZs with enclosed numbers in red circles and depict the bottom layer $Q$ valleys~($Q_b$ with green dots).  In particular, the $Q_b$ valley from the third BZ is close to the $K_t$ valley as indicated by the black rectangular boxes (Fig.1b).

The separation between the $K_t$ and $Q_b$ valleys further reduces in the quasicrystal when higher order Umklapp scatterings are considered shown in the left column of Fig.~1c.  
The top layer BZ is presented by the red hexagonal with $K_t$ valleys at the corners and the green dots represent the folded bottom layer's $Q$ valleys in the top layer BZ, i.e. $[Q_b]_{\text{BZt}}$. Here $[\boldsymbol{d}]_{\text{BZt}}$ means folding vector $\boldsymbol{d}$ into the top layer BZ. The distance $[K_t-Q_b]_{\text{BZt}}$ is smaller than 0.1 \AA$^{-1}$ when the third order Umklapp scattering is included (top panel) and can be arbitrarily close when higher-order Umklapp scatterings (e.g., up to the 15$^{th}$ order) are included (the bottom left panel of in Fig.~1c). In contrast, only a limited number of Umklapp scatterings are possible in the 21.8$^\circ$ moir\'e crystal due to its periodicity. $[K_t-Q_b]_{\text{BZ}_t}$ remains larger than 0.3\AA$^{-1}$ after either the 3$^{rd}$ order (top right) or  15$^{th}$ order (bottom right) Umklapp scatterings included in Fig.~1c.

When an E-field perpendicular to the 2D plane is applied, an energy difference between the two layers U is introduced allowing $K_t$ and $Q_b$ states to hybridize, opening 
an energy gap as illustrated in Fig.~1d. The calculated bands with a $U= 70$ meV are shown in Fig.~1e ~(details in Methods). The color bar corresponds to the wavefunction projection weight onto the top layer. The electrons in hybridized states still predominantly reside in the top layer with a smaller probability distribution in the bottom layer, sufficient to induce strong interlayer coupling.    
Crucially, $K_t$-$Q_b$ hybridization is only possible in the quasicrystal where Umklapp scattering processes connect the two momenta.  This calculation considers the band alignments between the K and Q valleys determined by our scanning tunneling microscopy and spectroscopy measurements, as exemplified by the constant current scanning tunneling spectroscopy shown in Fig.~1f~(more details in Methods and Fig.~S1).

The optical measurements are performed on the hBN encapsulated WSe$_2$ twisted bilayers embedded in dual-gated devices. We first examine the evolution of reflectivity spectra as a function of E-field at a constant hole doping density of -1.15$\times10^{12}$~cm$^{-2}$ in Fig.~2a, which corresponds to the black dotted diagonal line in Fig.~S2b. We choose this doping density because optical spectra are relatively simple so that we can focus on changes unique to quasicrystals. Additional data sets for doping-dependent spectra with different E-fields in Fig.~S3.

When an upward E-field is applied, three distinct regions are observed as indicated by horizontal dashed lines. In Region I, characterized by a near-zero E-field with energy difference between the layers U$\sim$ 0 meV, doped holes are distributed across both layers.~In Region II with $E\in [E_0,0.08]$ V~nm$^{-1}$, the E-field is sufficiently strong to move the doped holes to the top layer, where $E_0$ is the minimal E-field required to create the charge configuration (p,i) in the bilayer~\cite{xu2022tunable}~(see charge configurations in Fig.~S2b). Because doped holes are in the top layer, we assume that the positive trions ($T_t^+$) mostly form in the top layer and the neutral excitons are mostly found in the bottom layer ($X_b$) as labeled in Fig.~2a and illustrated in Fig.~2d. Notably, the exciton exhibits a slight blue shift with increasing electric field under finite doping, consistent with its intralayer nature~\cite{wang2018electrical}. For similar reasons, the two resonances in region II' with a reversed E-field direction are attributed to the bottom layer trion ($T_b^+$) and top layer neutral exciton ($X_t$).  The energy differences between trions and excitons in regions II and II' are likely due to coupling to the substrate. With increasing doping density, the oscillator strength shifts gradually from excitons to trions. 

The most drastic changes occur in region III, where the applied E-field shifts the energy between the layers sufficiently to hybridize states in the $K_{t}-Q_{b}$ valleys. The energy difference is calculated as $U = \boldsymbol{E} \cdot \boldsymbol{d}$, where $d$ represents the effective thickness of the bilayer, chosen to be 0.7 nm~\cite{he2014stacking,liu2014evolution}. The $K_{t}-Q_{b}$ valley hybridization opens an energy gap in the conduction band as predicted in Fig.~1e, leading to the hybridized exciton doublet $X_h$ (indicated by grey dots and arrows). The exciton doublet is most visible in the line cuts (top two curves) shown in Fig.~2b and Fig.~S4a with a broader range of electric fields. The electron wavefunction distributes in both layers as a consequence of the hybridization, giving rise to an out-of-plane component of the excitonic dipole moment. Thus, a red-shift of the exciton resonances $X_h$ with an increasing E-field becomes observable.

Another distinct change in the reflectivity spectra exceeding the critical E field is the disappearance of the trion state $T_t^+$, concurrent with the emergence of the hybrid exciton doublet $X_h$ in Region III. We attribute the trion disappearance to the predicted quasicrystal localization that only emerges with strong interlayer coupling~\cite{park2019emergent}. We calculated the inverse participation ratio (IPR) that characterizes how localized a given state is based on a simple model (more details in Methods). A larger IPR means a more localized state. The effective interlayer coupling strength and IPR are shown in Fig.~2c. These two quantities exhibit similar dependence on the interlayer energy difference U and both quantities have a dramatic increase at around U=60 meV primarily due to $K_t-Q_b$ hybridization. At higher U, the reduced coupling strength and IPR are not physically meaningful but are rather artifacts of the simple model~(only 3$^{rd}$ Umklapp scattering is considered). In addition, the simplified model we use cannot yield a full localization but rather a tendency to localize the hybrid electrons (more details in Methods).
Fig.~2(d,e) illustrates the layer distribution of doped holes, excitons and trions. In real space, excitons and doped holes bind to form trions (Fig.~2d). As layer hybridized electrons (green circles) 
become localized, the probability of forming trions reduces, and the oscillator strength shifts back to excitons. At much higher hole densities, trions may still form as we elaborate later.

We now compare the hole- and electron-doped quasicrystal in the presence of an upward E-field as illustrated in Fig.~3a-b. The different dopant distributions in real space and valleys lead to different behaviors of the trion resonance. The doped holes reside in the top layer
, and the Fermi level of the hole-doped quasicrystal is below the valence band maximum at the $K_t$ valley (Fig.~3a). When the E-field exceeds the critical value to cause $K_t-Q_b$ hybridization, layer hybridized electrons at the K valleys (green circles) and hybrid excitons become localized, leading to the disappearance of $T_t^+$.  In electron-doped quasicrystal, both doped electrons at $Q$ and excitons $X_b$ in the bottom layer do not participate in layer hybridization, and they are still delocalized. Thus, negatively charged trions ($T_b^-$) in bottom layer are still observable even when the E-field exceeds the critical value. Reflectivity spectra at a few selected applied E-fields for the quasicrystal device 2 (D$_2$) at a fixed electron doping density of $1.1\times 10^{12}$ cm$^{-2}$ are shown in Fig.~3c.  
Interlayer $K_t-Q_b$ hybridization still leads to the hybridized exciton doublet X$_h$ (grey and purple points) in Fig.~3c. With an increasing E-field, their out-of-plane dipole moment component leads to a red-shift. In addition, the oscillator strength shifts to the higher energy branch consistent with our calculation presented in Fig.~S8. While hybridized excitons still localize, we no longer observe any corresponding spectroscopy feature from $T_b^-$ that reside in the bottom layer.

The reflectivity spectra at a comparable hole doping density $-1.5\times 10^{12}$ cm$^{-2}$ in a moir\'e crystal provide another meaningful comparison. In Fig.~3d, the blue (red) dots mark the peak energies of the top-layer trion $T^{+}_{t}$ (bottom layer exciton $X_{b}$). Neither resonance exhibits abrupt changes (i.e. no exciton resonance splitting or trion disappearance) when the E-field is varied within the range of 0 - 0.13 $V$ nm$^{-1}$. Similar data for electron doped moir\'e crystal (shown in Figure.~S5) also displays no significant E-field-dependent changes. These experiments performed on moir\'e crystals support our hypothesis that the tunable exciton localization originates from the unique $K-Q$ coupling in quasicrystals. The localization mechanism is distinct from that caused by disorder (e.g., inhomogeneous strain or impurities), which are present in both types of bilayers with either commensurate or incommensurate atomic arrangements. 

Since only excitons consisting of layer hybridized electrons are localized, the trion resonance $T^{+}_{t}$ may still form if the doping density is sufficiently high. We take reflectivity spectra as a function of doping density when the applied E-field exceed the critical value shown in Fig.~4a. The disappearance of trion $T^{+}_{t}$ in hole-doped quasicrystals is only observed over a certain range of doping density. The trion resonance re-emerges when the doping level exceeds $-3.5\times 10^{12}~cm^{-2}$ under $0.1V nm^{-1}$ E-field. The probability of trion formation remains finite despite exciton localization or a somewhat screened E-field at higher doping density (See 2D diagram and linecuts in Fig.~S6). 

We compare the Zeeman shifts of hybrid exciton X$_h$  and bottom layer exciton X$_b$ to provide further evidence that they consist of electrons in different valleys. Magneto-optical reflectivity spectra shown in Fig.~4b are taken under a 0.1 V $nm^{-1}$ electric field and a doping density of $-1.5\times 10^{12}~cm^{-2}$, when both X$_h$ and X$_b$ are present. The g factor is calculated using the relationship~$\Delta$E =~g$\mu_B$B, where $\Delta$E is the energy difference between $\sigma^-$ and $\sigma^+$ excitations. Following a linear fit, we extract the g factors for X$_h$ and X$_b$ to be -7.9~$\pm0.7$ and -3.9~$\pm0.6$, respectively. In the simplest model, the exciton g factor is calculated by: g = -2$\times$(L$_v$-L$_c$), where L$_v$ (L$_c$) is the orbital angular momenta of valence (conduction) states of the hole (electron). Unlike the K valleys, the Q valleys in the conduction band possess significantly smaller electron orbital angular momentum~\cite{forste2020exciton}. A smaller L$_c$ of the X$_h$ due to $K-Q$ hybridization results in its larger negative g factor value than that of X$_b$, where both the electron and hole are in K valleys. This analysis confirms our interpretation that X$_h$ consists of hybridized electrons with its wavefunction predominantly in the top layer.

We discuss the subtleties of the localization mechanism in the quasicrystal and alternative explanations to our observations. Localization in quasicrystals is expected from single particle theories and does not require the inclusion of Coulomb interaction. This localization is robust to small variations in twist angles\cite{moon2019quasicrystalline}. While we cannot completely rule out that disorder-induced Anderson localization plays a role, the distinct behavior observed in quasicrystals and the periodic moir\'e crystal suggests that our observations likely originate from unique quasicrystal properties.~In our case, the two monolayers consisting of the WSe$_2$ quasicrystal are largely decoupled electronically at zero applied field. The field-induced $K-Q$ hybridization imposes a strong interlayer coupling. While the disappearance of the trion resonance clearly indicate spatial distribution of wavefunction is changed by the enhanced layer coupling, our optical experiments do not probe the wavefunction distribution directly.  
Due to the aperiodic ordering of atoms in quasicrystals, destructive interference in electron hopping occurs when electrons are subject to the potentials imposed by the other layer or superposition of incoherent single-particle states with different wavevectors may hinder the trion formation. This phenomenon is common in twisted bilayer quasicrystals with large interlayer couplings including twisted graphene  bilayers~\cite{moon2019quasicrystalline,park2019emergent,gonzalez2022localization}. 
Future experiments with sufficiently high spatial resolution may be able to directly image electrons or excitons and, thus, distinguishing complete localization from fractal wavefunction distribution theoretically predicted for quasicrystals~\cite{zhang2024critical}.

In conclusion, we identify unique field-dependent evolution in the optical spectra of WSe$_2$ dodecagonal quasicrystals due to their aperiodic atomic arrangements. These features are attributed to efficient higher-order Umklapp scatterings that bring the $K$ and $Q$ valleys in separate layers arbitrarily close in the momentum space. This unique $K-Q$ coupling is further enhanced when a modest perpendicular E-field aligns them energetically, leading to highly unusual changes in the optical spectra of a hole-doped  WSe$_2$ quasicrystal. Not only hybrid exciton doublet forms due to an energy gap from $K-Q$ hybridization, but also the trion resonance disappears, which we attribute to hybrid exciton localization from the enhanced interlayer coupling. The interpretation that the hybrid exciton consists of mixed $K-Q$ electrons is supported by its different g-factor from an intralayer exciton. We show these features are unique to quasicrystals by comparing them with those spectra measured in a moir\'e crystal. Many intriguing phenomena predicted in quasicrystals remain to be explored, including 
fractal spectra, and energy-level statistics~\cite{bandres2016topological,jagannathan2021fibonacci,liu2021multichannel}. Our studies also provide critical guidance to future experiments on increasingly more complex vdW heterostructures. One example is double moir\'e structures with incommensurate stacking between the two moir\'e superlattices~\cite{uri2023superconductivity}. One can also envision constructing vdW quasicrystals with diverse functional components, capitalizing on the potential of hybridizing orbitals at distinct valleys--a novel approach to valley engineering.

\section*{Methods}

\subsection{Device fabrication and characterization:} 

We incorporate twisted WSe$_2$ bilayers assembled using the "tear and stack" technique, along with hBN encapsulation layers, into dual-gate devices. Quasicrystal data presented in the main text are taken from two devices ($D_1$ and $D_2$). These devices feature a few-layer graphite top gate (TG) and a metallic back gate (BG). These components allow for independent tuning of carrier density and electric field. For the back gate, we employ a deposition of 3 nm of Cr and 3 nm of Pt. This combination ensures good electrical conductivity while maintaining reasonable thickness. The patterned electrodes are deposited using a layering of 10 nm Cr, 20 nm Au, and 10 nm Pt. Charge densities n are calculated using a parallel capacitor model:
\begin{equation}
    n=\frac{\epsilon_r \epsilon_0(V_t-V_{t0})/d_{top}+\epsilon_r \epsilon_0(V_b-V_{b0})/d_{bottom}}{e}.
\end{equation}
Here $\epsilon_r$ is the relative permittivity of hBN ($\sim$4.2~\cite{regan2020mott}) and $V_{t}$ and $V_{b}$ are top and bottom gate voltages. $V_{t0}$ and $V_{b0}$ are offsets that start to dope the twisted bilayer which is read from dual-gate mapping as shown in Fig.~S2b. The thickness of the top and bottom hBN layers of D$_1$~(i.e.,~$d_{top}$ and $d_{bottom}$) in D$_1$ (D$_2$) are found to be 39~nm (38~nm) and 40~nm (41~nm) respectively from atomic force microscope measurements. We calculate the applied out-of-plane electric field using the formula:
\begin{equation}
    E_z=(C_{bottom}V_{b}-C_{Top}V_{t})/2\epsilon_{WSe_2},
\end{equation}
where $C_{bottom}(C_{Top})=\epsilon/d_{Bottom}(d_{Top})$ is the geometric capacitance of the bottom(top) hBN. We use $\epsilon_{WSe_2}$=7.8~\cite{movva2018tunable}.
\subsection{Optical measurements:} 

Optical measurements are carried out at 4.4 K in a home-built confocal microscope using an objective with a numerical aperture of 0.5. A halogen lamp is used as the light source. We define the differential reflectance as $\Delta R/R_0$ where $R_0$ is the background measured from a region without the WSe$_2$ flake. The reflectance spectra are fitted with a multi-Lorentzian function with a phase shift $\alpha$ that depends on the energy and gate voltage\cite{smolenski2019interaction,shimazaki2020strongly}.~For each resonance,
\begin{equation}
    \Delta R/R_0=\frac{A cos(\alpha) \gamma^2/2}{(E-E_0)^2+\gamma^2/4}+\frac{Asin(\alpha)\gamma(E-E_0)}{(E-E_0)^2+\gamma^2/4}+c.
\end{equation}
To emphasize the resonances, we sometime take the derivative of the differential reflectance with respect to energy, i.e., $d(\Delta R/R_0)/d(E)$.
\subsection{STM and STS measurements}
STM and STS measurements were conducted at 4.3$K$ in the STM chamber, with a base pressure of about $10^{-11}$ torr. The W tip was prepared by electrochemical etching and then cleaned by in situ electron-beam heating. STM dI/dV spectra were measured using a standard lock-in technique with a modulation frequency of 758~Hz. Two different modes of STS were simultaneously used: (1) the conventional constant-height STS and (2) the constant-current STS. The assignment of Q and K valleys in the conduction band follows the method developed by Zhang et al \cite{zhang2015probing}, by inspecting the behavior of the tip-sample distance (Z) and the tunneling decay constant ($\kappa$) as a function of sample bias (shown in Fig.~S1d-e) during the constant-current STS scan. The Z-V identifies the thresholds for different valleys (Fig.~S1d). Meanwhile, the decay constant $\kappa$ for the Q valley is smaller than that for the K-valley due to having a smaller $|q|$ in the BZ than that for the K-valley (Fig. S1e).  

\subsection{Model calculation}

The distance between interlayer $K$ and $Q$ points in reciprocal space can be generally written as $\boldsymbol{d}_{K_t-Q_b}=(K_t+\boldsymbol{G}_t)-(Q_b+\boldsymbol{G}_b)$, where $K_t$ and $Q_b$ are the $K$ and $Q$ points within their first Brillouin zone, and $\boldsymbol{G}_{t(b)}$ is the reciprocal lattice vectors of the top (bottom) layer. One way to show this distance is to fold it into the top layer Brillouin zone ($\text{BZ}_t$): $[\boldsymbol{d}_{K_t-Q_b}]_{\text{BZ}_t} = [(K_t+\boldsymbol{G}_t)-
(Q_b+\boldsymbol{G}_b)]_{\text{BZ}_t} = K_t-[Q_b+\boldsymbol{G}_b]_{\text{BZ}_t}$. Here $[\boldsymbol{d}]_{\text{BZ}}$ means folding the vector $\boldsymbol{d}$ into the corresponding Brillouin zone. For the quasicrystal, as shown in Fig.~1(b,c), there are regions in the BZ (blue rectangles) where the distance between the $K_t$ and $Q_b$ valleys is smaller than 0.1 \AA$^{-1}$ when the $3^{rd}$ order Umklapp scattering is taken into account.
At the $3^{rd}$ order, we see in total 12 different $Q_b$ points near the $K_t/K'_t$ points. If we focus on one single valley of the top layer, there should be 6 different $Q_b$ points surrounding it. We approximate the band structure near the $K$ and $Q$ points using simply the parabolas with different effective masses, we should have the relative band alignment illustrated in the left panel of Fig.~1(d), in which a proper displacement field will couple the $K_t$ point of the top layer with the bottom bands (near but not at the $Q_b$ points). For 21.8$^\circ$ moir\'e crystal however, the $Q$ valleys are at $|\Gamma-Q_b| =0.57~|\Gamma-K_b|$~\cite{debnath2022tuning}. The smallest momentum mismatch between $K_t$ and $Q_b$ is larger than 0.3 $A^{-1}$.

We can now write a simple effective Hamiltonian to describe the $K-Q$ coupling near the third order:
\begin{equation}
    H^{K-Q}_{\text{eff}} = 
    \begin{pmatrix}
        E_{Q_b}(\boldsymbol{k}_1)&  &  &  & & & t_q \\
         &E_{Q_b}(\boldsymbol{k}_2)&  &  & & & t_q \\
         &   &E_{Q_b}(\boldsymbol{k}_3) &  & & & t_q \\
         &   &  &E_{Q_b}(\boldsymbol{k}_4) & & & t_q \\
         &   &  & &E_{Q_b}(\boldsymbol{k}_5)& & t_q \\
         &   &  & & &E_{Q_b}(\boldsymbol{k}_6)& t_q \\
        t^*_q & t^*_q & t^*_q&  t^*_q& t^*_q&  t^*_q & E_{K_t}(\boldsymbol{k})+U
    \end{pmatrix},
\end{equation}
where $\boldsymbol{k}_i = \boldsymbol{k}-(Q_{bi}-K_t)$ with $Q_{bi}$ the surrounding $Q$ points, $E_{K_t}(\boldsymbol{k}) = \boldsymbol{k}^2/m_{K_t}$ and $E_{Q_b}(\boldsymbol{k}) = \boldsymbol{k}^2/m_{Q_b} + \Delta$ are the parabolic band dispersion near the $K$ and $Q$ points, respectively, with effective masses $m_{K_t}$ and $m_{Q_b}$ and the off-set energy $\Delta$. We choose the off-set energy to be 110 meV, extracted from STM result (Figure.S1) and $m_{Q_b}$ is 0.7 $m_0$ where $m_0$ is the bare electron mass.  Here $t_q$ is the interlayer transfer integral (tunneling amplitude). Those $t_q$'s should ideally have phase differences that are related by symmetries like the $C_3$ rotation symmetry since they are different $K-Q$ couplings of the same order. However, here it is sufficient to use only their amplitudes to demonstrate our point. We explicitly shift the conduction band $K$ valley of the top layer by an energy $U$ to account for the effect of the displacement field. By diagonalizing the Hamiltonian exactly, we obtain the band structure near the top layer $K$ point as shown in Extended Fig.~6(a-c) with different displacement fields.

The calculation results show interlayer $K$-$Q$ coupling can be achieved in a quasicrystal sample (with 30$^\circ$ twist) with a moderate interlayer energy difference $U=60$ meV. The $U$ required for the moir\'e crystal sample (with 21.8$^\circ$ twist) is huge, around 1 eV, given that the minimal distance $d_{K_t-Q_b}$ in the moir\'e crystal sample is much larger than that in the quasicrystal (see Fig. 1 in the main text), which far exceed the range experimentally achievable.

\subsection{Inverse participation ratio and Localization}
To see the indication of localization from this largely simplified model, we can compute the inverse participation ratio (IPR) of the wavefunction near the $K_t$ point of the conduction band. The IPR of a state $ \psi(x,y)$ is defined in the following way
\begin{equation}
    \text{IPR}_{\psi} = \frac{\sum_{x,y}|\psi(x,y)|^4}{\left[ \sum_{x,y}|\psi(x,y)|^2 \right]^2}
\end{equation}
which is a dimensionless quantity. Normally, one can drop the denominator if the normalization of the state $\psi(x,y)$ is already done. The IPR is maximized to 1 by a fully localized state and minimized to $1/N$ (with $N$ the system size, i.e., the number of grid points after discretizing the space) by a fully extended (delocalized) state. Given the simplified model, under the $K-Q$ coupling, we do not expect a full localization due to the $K-Q$ coupling but a tendency to localize the $K_t$ electron (namely larger IPR). To further illustrate our point, we can write the wavefunction at $K_t$ and $Q_{bi}$ points in the general form:
\begin{equation}
    \psi_{K_t}(\boldsymbol{r}) = e^{i\boldsymbol{K}_t\cdot\boldsymbol{r}}\sum_{\boldsymbol{G}_t}f^{K_t}_{G_t}e^{i\boldsymbol{G}_t\cdot\boldsymbol{r}}, \qquad \psi_{Q_{bi}}(\boldsymbol{r}) = e^{i\boldsymbol{Q}_{bi}\cdot\boldsymbol{r}}\sum_{\boldsymbol{G}_b}f^{Q_{bi}}_{G_b}e^{i\boldsymbol{G}_b\cdot\boldsymbol{r}}
\end{equation}
It is instructive to use a simple Gaussian decay profile to approximate the Fourier components $f^{K_t}_{G_t}\sim e^{-|\boldsymbol{K}_t+\boldsymbol{G}_t|^2/\xi_t} $ and $f^{Q_{bi}}_{G_b} \sim e^{-|\boldsymbol{Q}_{bi}+\boldsymbol{G}_b|^2/\xi_b}$ as we shall explain later the essence of the localization in the quasicrystal. Those wavefunctions serve as the basis for Hamiltonian in Eq.~(1). Then, one can readily obtain the IPR for states near the $K_t$ point under the displacement field, i.e., the $K-Q$ couplings.

The IPR and the effective coupling strength are both shown in Fig.~2d, which shows a strong correspondence between the $K-Q$ coupling and the localization. Such localization is actually expected provided that the state near the $K_t$ point is now a hybridization between the plane waves with wavevectors $\boldsymbol{K}_t+\boldsymbol{G}_t$ and $\boldsymbol{Q}_{bi}+\boldsymbol{G}_b$. These wavevectors are incommensurately related to each other leading to an incoherent supposition of waves forming a localized wavepacket. It is then understandable that stronger $K-Q$ couplings necessarily give more localized states, namely larger IPR.

\section*{References}
\bibliography{Tunable_Localization-v1}

\section*{Data availability}
The data that support the plots within this paper and other findings of this study are available from the corresponding authors upon reasonable request. Source data are provided in this paper.

\section*{Acknowledgments}
This study is primarily supported by the National Science Foundation through the Center for Dynamics and Control of Materials: an NSF MRSEC under Cooperative Agreement No. DMR-1720595 and DMR-2308817. (Z.L., Q.G., Y.L., D.S.K., C.S., E.K., and X. Li).  Y. N. and X. Liu gratefully acknowledge the Department of Energy, Office of Basic Energy Sciences under grant DE-SC0019398 for device fabrication. H.A. is supported by the Welch Foundation Chair F-0014  and NSF ECCS-2130552. K.W. and T.T. acknowledge support from the JSPS KAKENHI (Grant Numbers 20H00354 and 23H02052) and World Premier International Research Center Initiative (WPI), MEXT, Japan.

\section*{Author contributions}
Z.L., K.S., E.K., and X.Li conceived the project. Z.L. fabricated the samples with assistance from Y.N., X.Liu, and M. M. Z.L. analyzed the data with contributions from Y.L.,~D.S.K, H.A., and X.Liu. K.W and T.T synthesized the hBN bulk crystals. Q.G. and E.K. proposed the theoretical model. Z.L., Q.G., E.K., and X.Li wrote the first draft of the manuscript. All authors contributed to discussions.

\section*{Competing interests}
The authors declare no competing interests.

\clearpage
\renewcommand{\baselinestretch}{1}

\clearpage
\newpage
\section*{Figures}

\begin{figure}[htp]
 \includegraphics[width=15cm]
{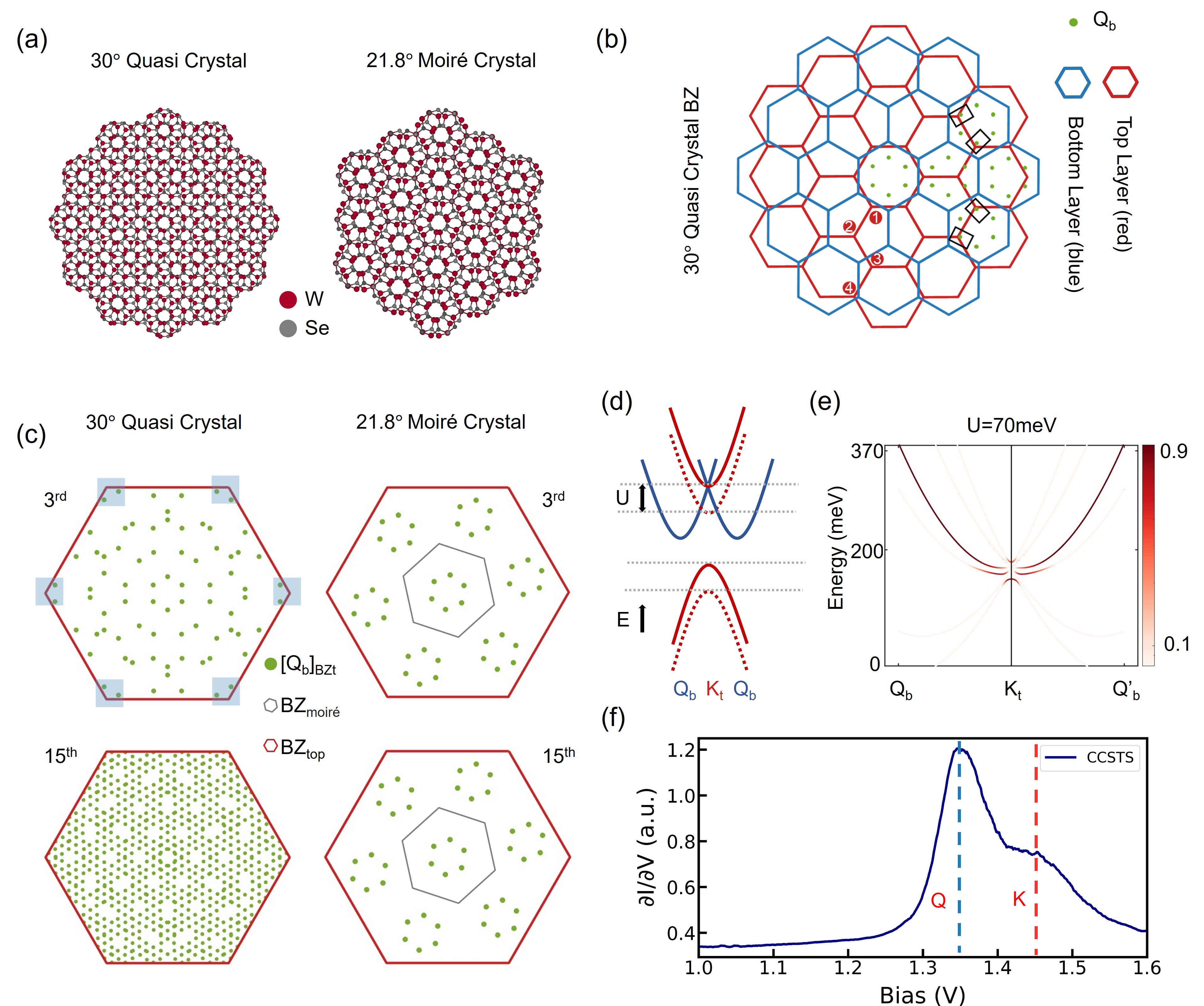}
    \caption{\textbf{Inter valley K-Q coupling in WSe$_2$  quasicrystals vs. moir\'e crystals .}
    \textbf{(a)} Real space lattices of the 30$^\circ$ quasicrystal and 21.8$^\circ$ moir\'e crystal. Red (grey) dots represent tungsten (selenium) atoms.
     \textbf{(b)}   Extended Brillouin zone (BZ)  for the top (red) and bottom (blue) layers. Green dots indicate the Q valleys of the bottom layer at the conduction band minimum. In this extended zone picture, the distance to the $\Gamma$ point measures the order of interlayer Umklapp processes as indicated by the enclosed numbers. The black boxes highlight the $K-Q$ separation considering the third-order Umklapp scattering.
    \textbf{(c)} Simulated Umklapp scattering where green dots represent the folded bottom layer's $Q$ valleys in top layer BZ and the corners of red hexagonal are $K_t$ valleys. The top (bottom) row presents calculations after the 3$^\text{rd}$ order (15$^\text{th}$ order) Umklapp scatterings are taken into account.~The blue-shaded squares highlight regions with the closest $K-Q$ distance. The distance between $K_t$ and $Q_b$ is arbitrarily small (finite) after higher-order Umklapp scatterings are included in quasicrystals (moir\'e crystals). The grey hexagonals in the right panels represent the moir\'e crystal BZ.
    \textbf{(d)} Illustration of the interlayer coupling between top-layer K (red solid line) and bottom-layer Q valleys (blue lines) in the presence of an electric field in quasicrystals. 
    \textbf{(e)} Calculated hybrid $K-Q$ states at an interlayer energy difference U of 70 meV. 
    The color scheme indicates the wavefunction projection weight onto the top layer.
     \textbf{(f)} The conduction band Q~(K) valleys, marked by blue (red) dashed lines, are measured via constant current scanning tunneling spectroscopy ($\partial I/\partial V$).
    }  
    
    \label{fig:fig1}
\end{figure}

\newpage

\begin{figure}[htp]
 \includegraphics[width=16cm]{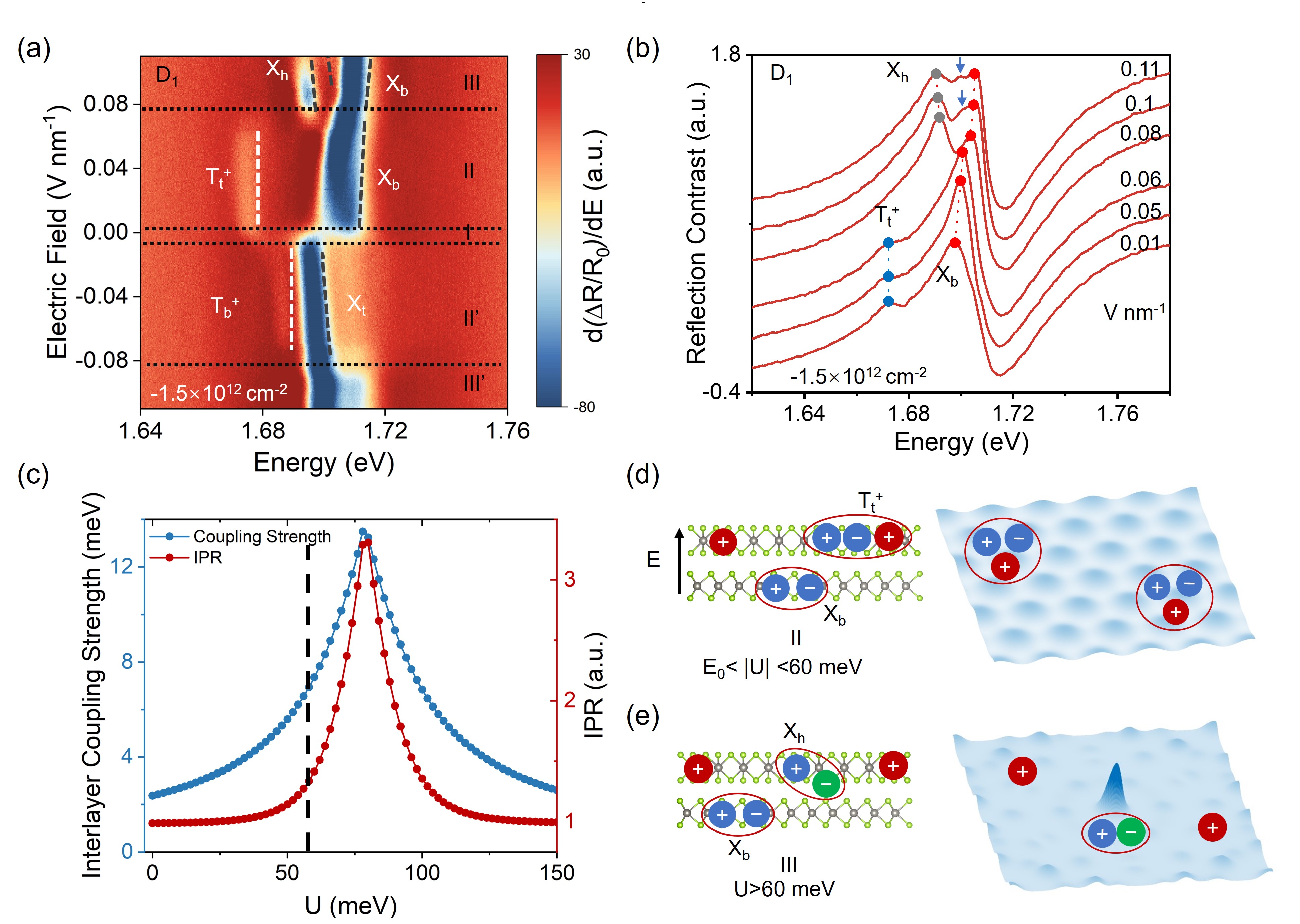}
    \caption{\textbf{Evolution of excitonic resonances with inter valley $K-Q$ coupling tunable by E-field.}
    \textbf{(a)} The reflectivity spectra of quasicrystal device 1 (D$_1$) plotted as a function of the electric field at a constant hole doping level of -1.5$\times10^{12}$ cm$^{-2}$. Three regions separated by horizontal black dashed lines represent increasing interlayer coupling tuned by the electric field. The white (black) dashed lines highlight intralayer trions and excitons. Hybrid exciton doublet and trion disappearance are observed in Region III.
    \textbf{(b)} 
    Reflectivity linecuts at various electric fields from panel (a). The blue, grey, and red dots label top layer trion ($T^{+}_{t}$), hybrid exciton ($X_{h}$), and bottom layer exciton ($X_{b}$). The arrows in the top two linecuts indicate hybrid exciton doublet due to $K_t-Q_b$ hybridization. 
    \textbf{(c)}
    Calculated effective interlayer $K-Q$ coupling strength and inverse participation ratio (IPR). The coupling strength and IPR increase when interlayer $K-Q$ hybridization occurs, indicating the trend of localization with enhanced interlayer coupling.
    \textbf{(d,e)} Left panels: layer distribution of doped holes (red circles), excitons, and trions for regions II and III. Right panels: illustrations of in-plane distribution of trions and localized excitons separated from the doped holes. Green circles represent hybridized electrons. 
    }
    \label{fig:fig2}    
\end{figure}

\clearpage

\newpage

\begin{figure}[htp]
 \includegraphics[width=16.5cm]{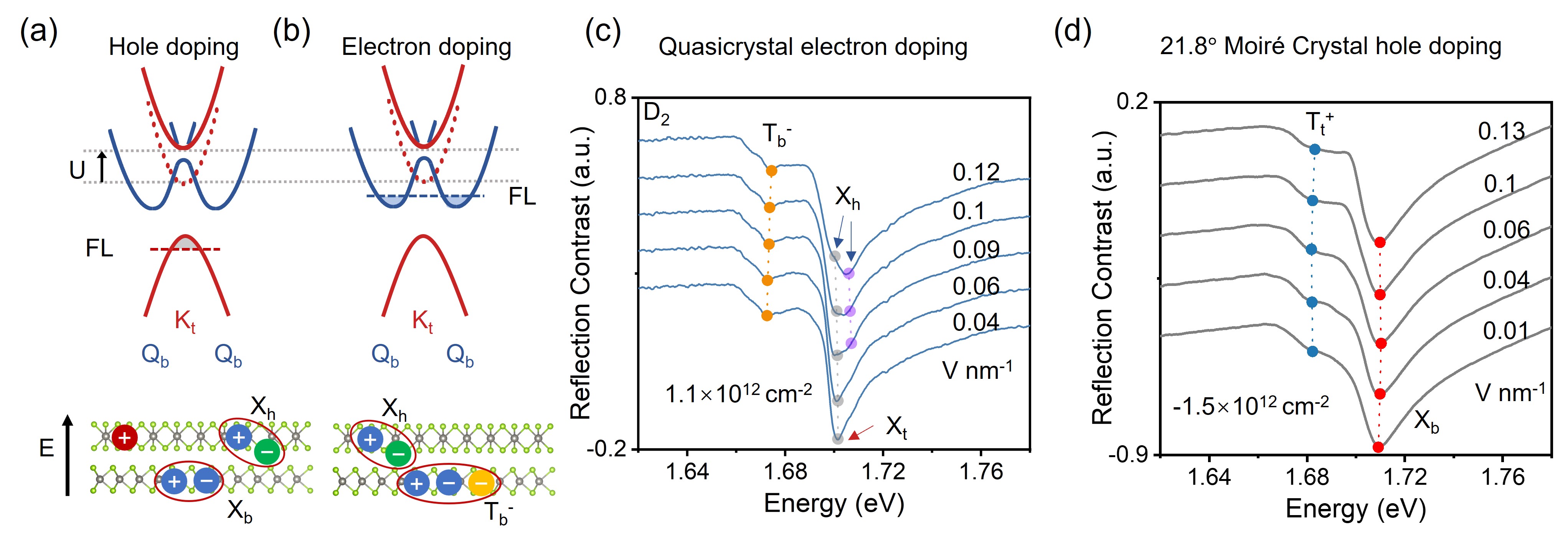}
    \caption{\textbf{ Optical reflectivity spectra from electron doped quasicrystal and hole doped moir\'e crystal.} 
    \textbf{(a,b)} Fermi level (FL), band alignments as well as layer distribution of dopants, excitons, and trions in hole- and electron-doped WSe$_2$ quasicrystals. Top panels: above a critical E-field, the bottom layer Q-valley (blue solid curves) hybridizes with the top layer K-valley (red solid curves). The red dashed curve indicates the K-valley without the interlayer energy difference (U).~The doped holes and electrons do not participate in layer hybridization, and they are delocalized. Bottom panels: layer distribution of doped holes (red circles), doped electrons (yellow circles), excitons, and trions. Only excitons consisting of hybridized electrons (green circles) are localized. An upward electric field moves the holes (electrons) to the top (bottom) layer.
     \textbf{(c)}
     Optical reflectivity spectra from the quasicrystal $D_2$ at several E-fields with  1.1$\times 10^{12}$~cm$^{-2}$ electron doping density, respectively. Hybrid exciton doublet forms above the critical field of 0.08 V nm$^{-1}$.
    \textbf{(d)} Optical reflectivity spectra at several E-fields taken from the $21.8^\circ$ moir\'e crystal at hole doping density of -1.5$\times10^{12}$~cm$^{-2}$.
    No abrupt changes are observed with an increasing E-field in the moir\'e crystal. 
    }
    \label{fig:fig3}    
\end{figure}
 
\clearpage

\begin{figure}[htp]
 \includegraphics[width=16cm]{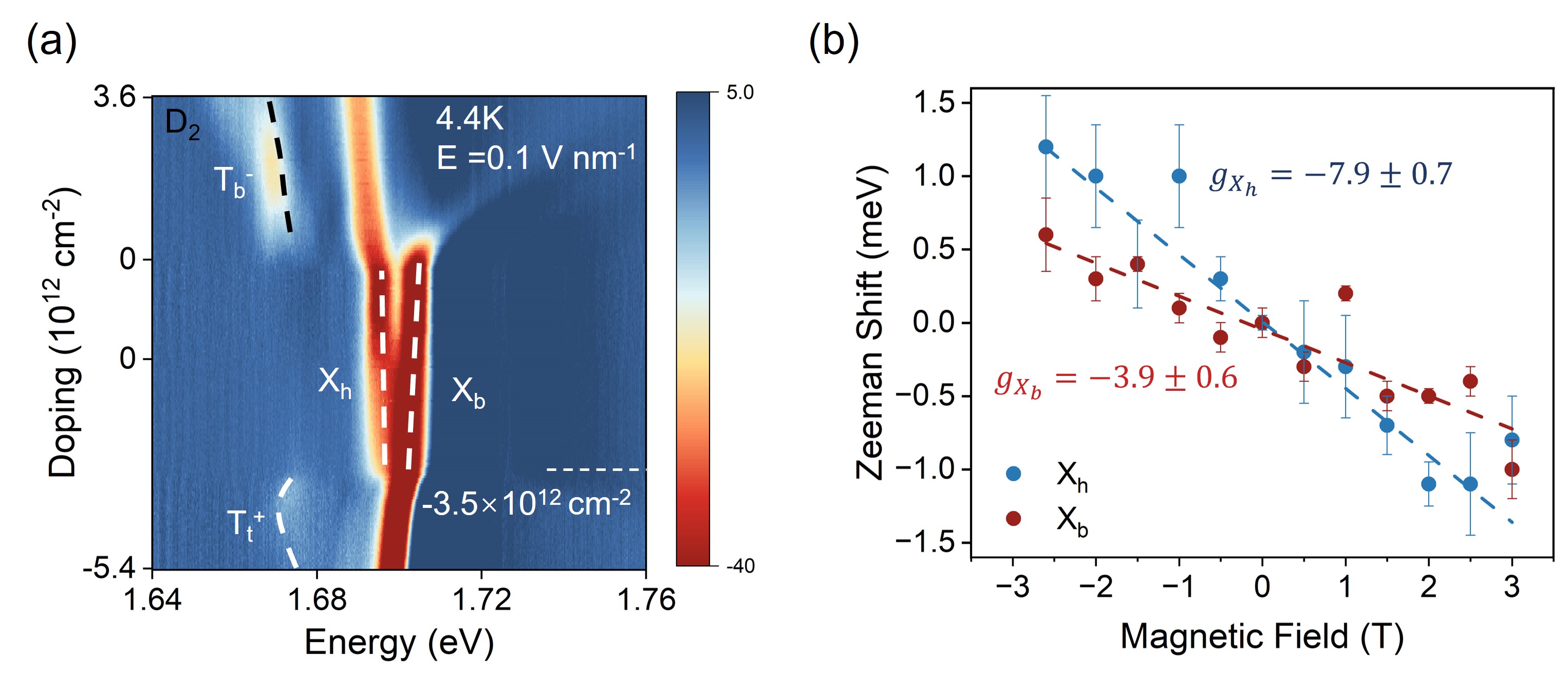}
    \caption{\textbf{ Doping density-dependent reflectivity spectra and Zeeman shifts in magnetic fields.}
    \textbf{(a)} Reflectivity spectra of quasicrystal D$_2$ as a function of doping with a constant field E=0.1 V~$nm^{-1}$. Two resonances, identified as hybrid exciton X$_h$ and the bottom layer exciton X$_b$, are observed in intrinsic regime and low doping density of holes. A positive trion signal T$_t^{+}$ re-emerges when the doping density exceeds $-3.5\times 10^{12}~cm^{-2}$.
    \textbf{(b)} Zeeman shift of X$_h$ and X$_b$ as a function of an applied magnetic field. Dashed lines are linear fits. 
    }
    \label{fig:fig4}    
\end{figure}

\clearpage
\setcounter{figure}{0}
\setcounter{table}{0}

\newpage
\section*{Extended Data}

\makeatletter

\renewcommand \thefigure{S\arabic{figure}}

\renewcommand{\baselinestretch}{1}

\begin{figure}[h]
 \includegraphics[width=16.5cm]{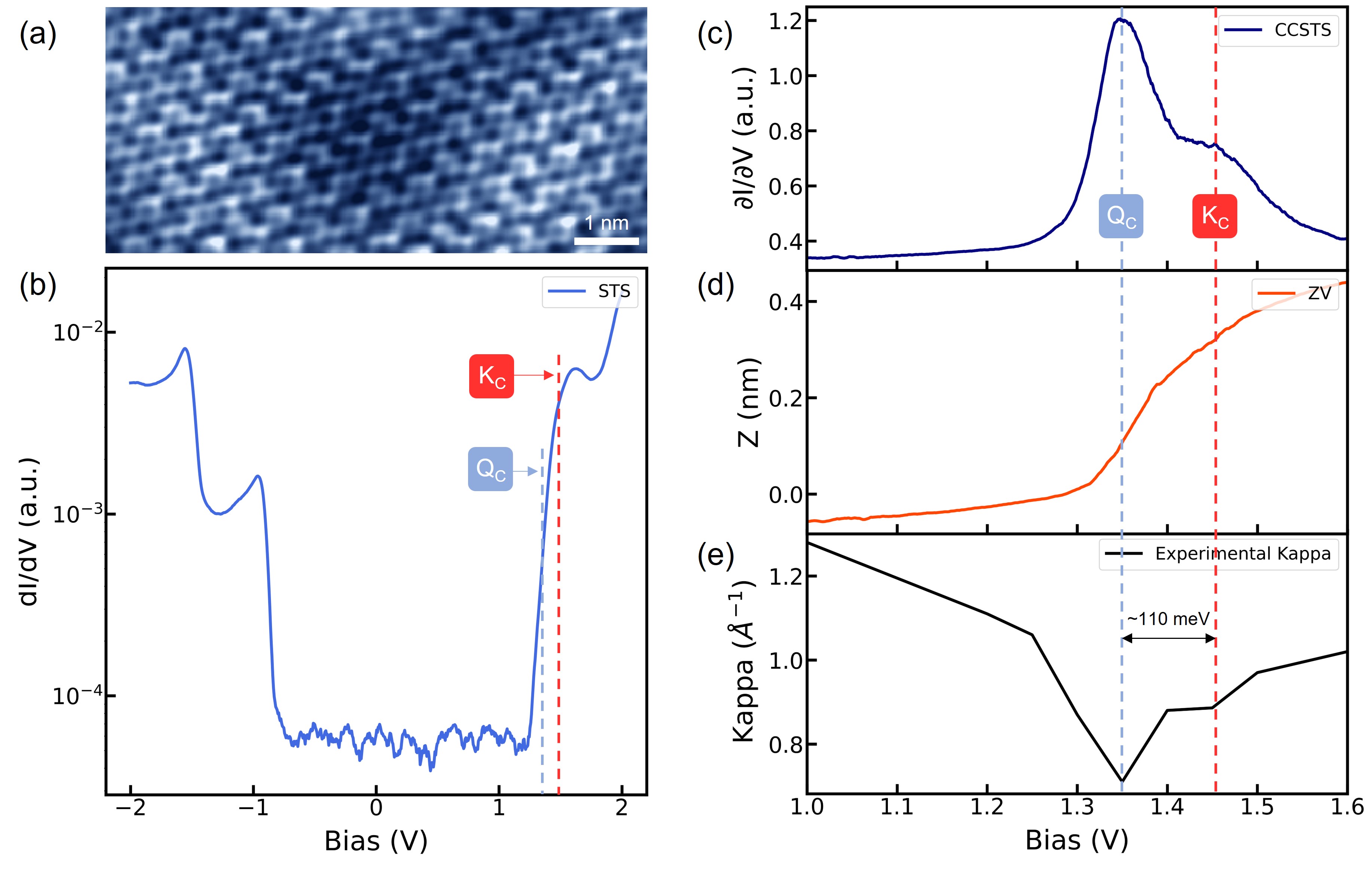}
  \caption{\textbf{Scanning tunneling spectroscopy measurement of the quasicrystal conduction bands at zero E-field at 4~K. } 
  \textbf{(a)} Typical scanning tunneling microscopy  topography image of 30° WSe2 bilayer. ($V_{bias}$ = -0.5~V, I = 30~pA) 
  \textbf{(b)} The spatially averaged scanning tunneling spectroscopy (STS) taken from a pristine region. The red (blue) dashed line indicates the energy position of the lowest K (second lowest Q) valley in the conduction band($V_{bias}$ = -2.0V, I = 100~pA)  
  \textbf{(c-e)} Variable-Z tunneling spectroscopy spectra measured at the same region. (c) The constant current STS ($\partial I/\partial V$), (d) tip sample distance (Z) versus bias voltage measurement (Z(V)) and (e) the experimental tunneling decay constant ($\kappa$). These measurements complement those in panel b to accurately identify the energies of different valleys. From these measurements, we conclude that K valley is higher in energy than Q valley by $\sim$ 110 meV, a value taken into account in our calculation.  
     }

    \label{fig:fig5}    
\end{figure}

\begin{figure}[ht]
 \includegraphics[width=16.5cm]{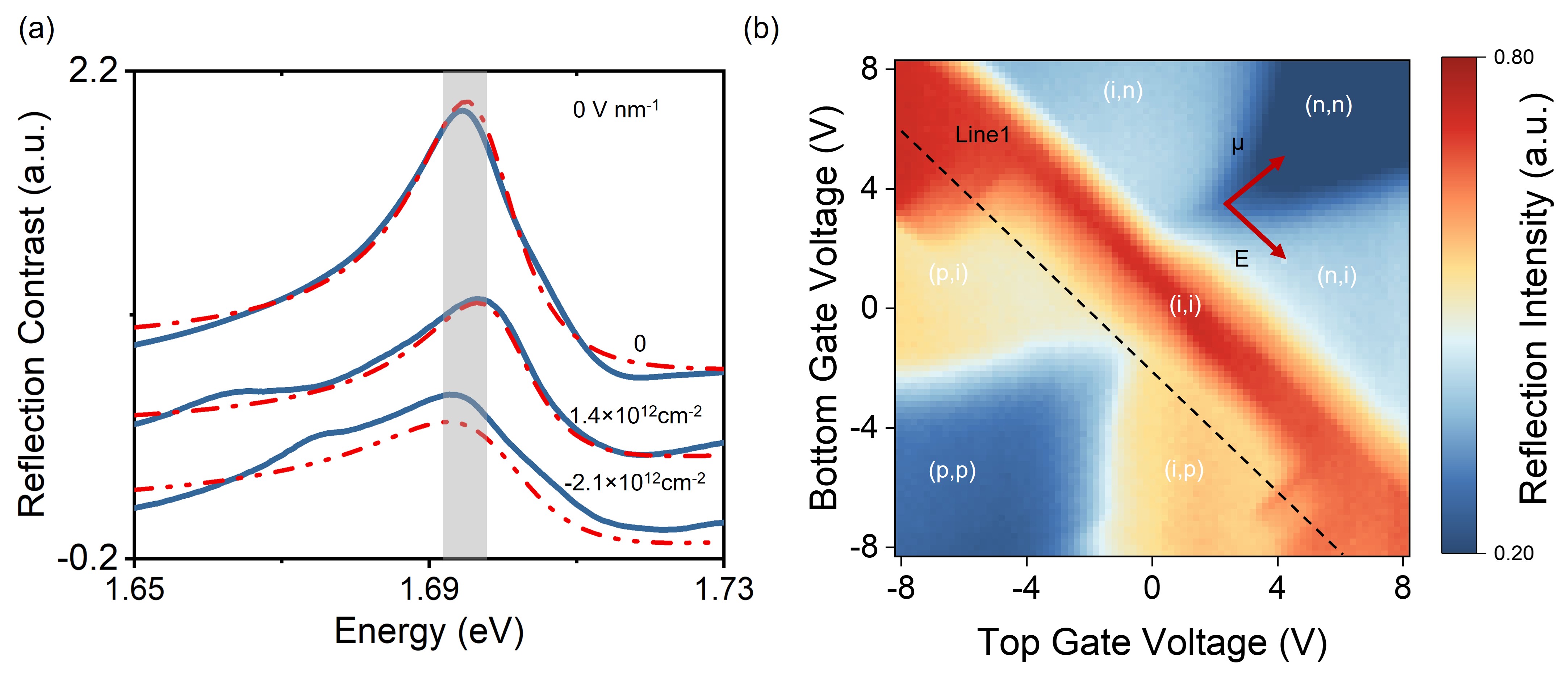}
 \caption{\textbf{Reflectivity spectra of quasicrystal D$_1$ and 2D map of 1S exciton intensity.} 
\textbf{(a)}Typical reflectivity spectra from the quasicrystal with no doping (top), with electron (middle), and with hole doping (bottom). Red dashed lines are fitting results of 1s exciton resonance. We use Lorentzian function with a phase shift as the fitting function. The grey vertical strip indicated the energy range used for 1s exciton intensity map in Fig.~S2b.
\textbf{(b)} A 2D reflectivity map as a function of top and bottom gates identifies the conditions under which the top (bottom) layer exhibits intrinsic (i), electron-doped (n), or hole-doped (p) as indicated by the first (second) index in the parenthesis. The two diagonal directions, marked by red arrows, allow independent control of doping density and an electric field to be perpendicular to the 2D layers.~The color bar represents the 1S exciton intensity in reflectivity spectra. 
    }
    \label{fig:fig7}    
\end{figure}

\begin{figure}[ht]
 \includegraphics[width=16.5cm]{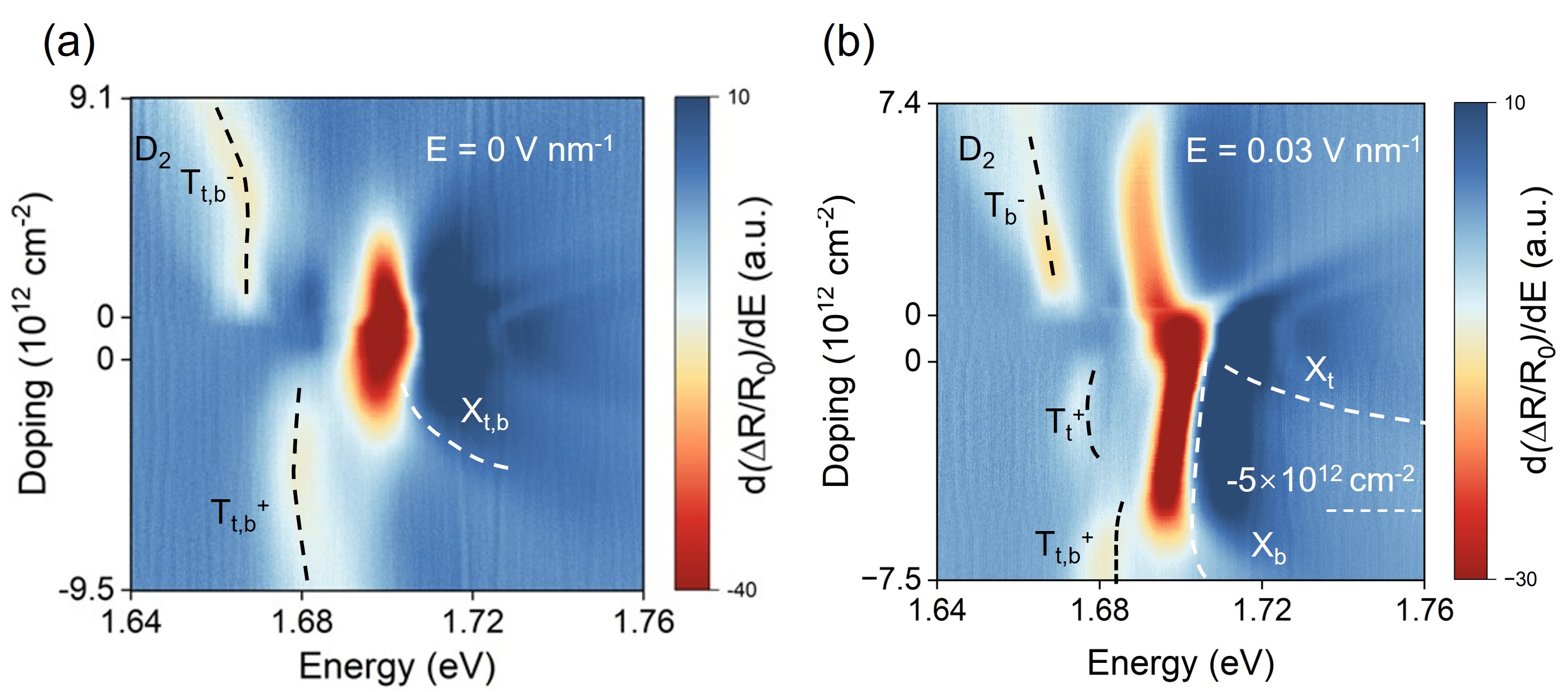}
 \caption{\textbf{Reflectivity spectra of quasicrystal D$_2$ as a function of doping with different E-fields.} 
  \textbf{(a)}~The reflection spectra are plotted as a function of doping density in the absence of an electric field.  Negative trion $T^{-}_{t,b}$ and positive trion $T^{+}_{t,b}$ are marked by black dashed lines. The blue-shifted exciton $X_{t,b}$ is marked as a white dashed line. The trion signals come from both layers.
  \textbf{(b)}~Reflection spectra are displayed as a function of doping density at a constant electric field of E=0.03 V nm$^{-1}$ which below the critical hybridization E field~($E_c$). Negative trion $T^{-}_b$ and positive trion $T^{+}_t$ are marked by black dashed lines. The blue-shifted exciton $X_t$ is marked by the white dashed line. The doped holes reside in the top layer below doping concentration $-5\times 10^{12}cm^{-2}$ and in both layers above it. The positive trion signal from both layers is marked as $T^{+}_{t,b}$.
    }
    \label{fig:fig6}    
\end{figure}

\begin{figure}[h]
 \includegraphics[width=16cm]{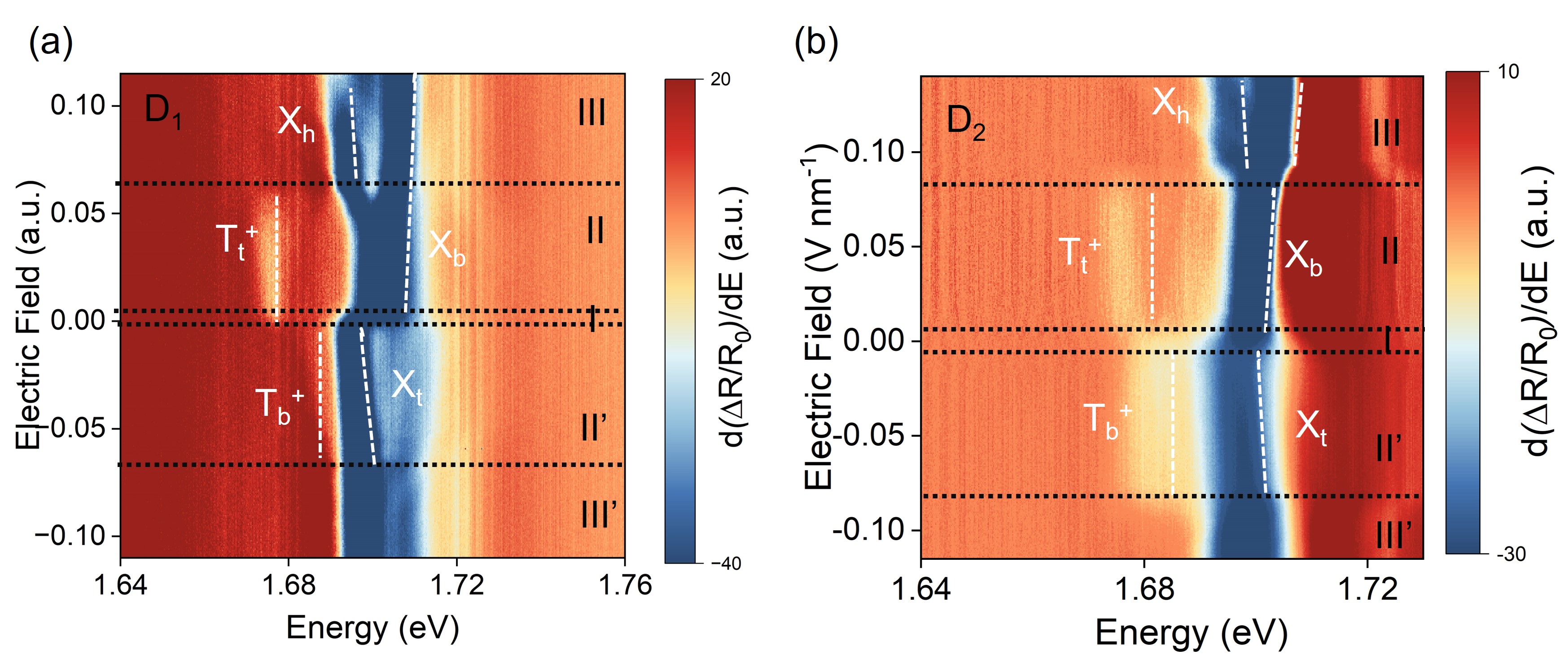}
\caption{\textbf{The reflectivity spectra of D$_1$ and D$_2$ plotted as a function of electric field at a constant hole doping levels.} 
\textbf{(a)} Larger range of electric field dependence of D$_1$ at a constant hole doping -1.15$\times10^{12}$ cm$^{-2}$. Hybrid exciton doublet at higher electric field is more visible in region III.
\textbf{(b)}
Reflectivity spectra from D$_2$ with electric field at hole doping -1.15$\times10^{12}$ cm$^{-2}$. Disappearance of the trion and hybrid exciton X$_h$ are observed, similar in Fig.~2a in the main text.}
    \label{fig:fig12}    
    
\end{figure}

\begin{figure}[ht]
 \includegraphics[width=10cm]{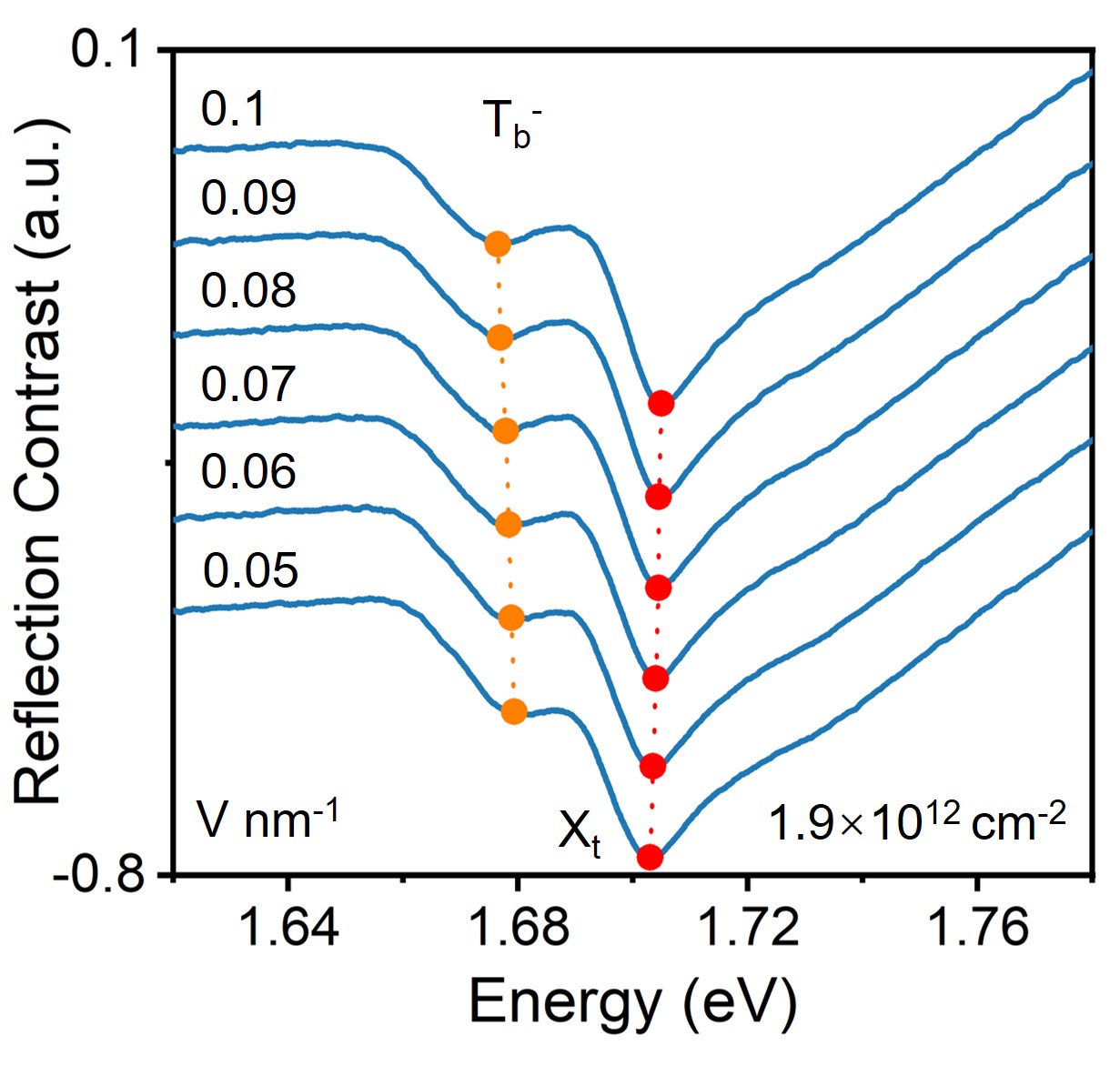}
\caption{\textbf{Reflectivity spectra 21.8$^\circ$ moir\'e crystal at a constant electron doping.} 
 Reflectivity spectra at a few E-fields at a constant electron doping 1.9$\times10^{12}$cm$^{-2}$. Orange dots show negative trion resonance $T_{b}^{-}$ and red dots show exciton resonance $X_{t}$. No abrupt changes are observed.
}
    \label{fig:fig10}    
\end{figure}

\begin{figure}[ht]
 \includegraphics[width=16.5cm]{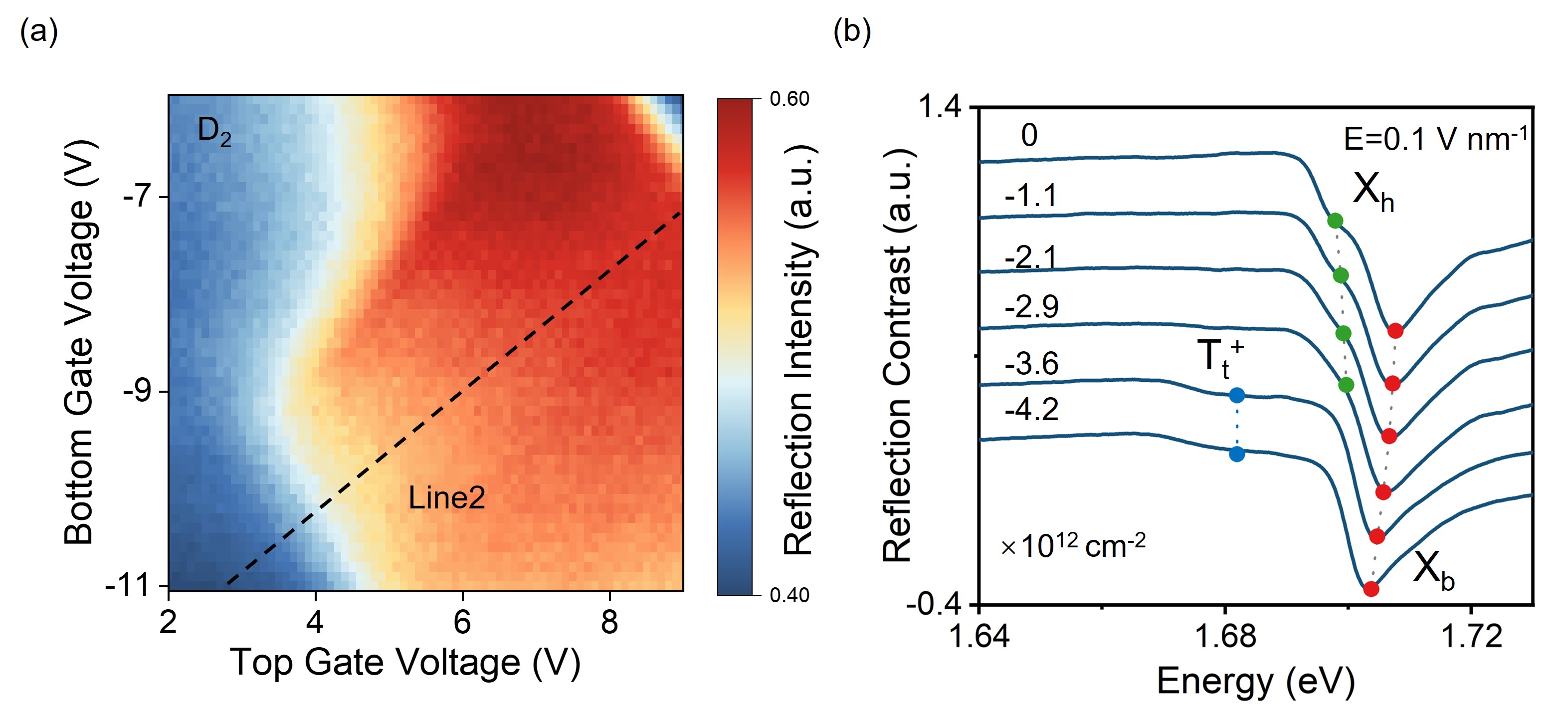}
  \caption{\textbf{Intensity map of 1s exciton in the high electric field region and reflectivity spectra of the hole-doped quasicrystal.}
   \textbf{(a)}
  The dashed black line shows the doping-dependent reflectivity spectra in Fig.~4a at a constant  E-field $0.1 V~nm^{-1}$. Along this line, the doping density is varied from -5.4 to 3.6$\times 10^{12}~cm^{-2}$. In the region with red (blue) color bar, minimal (strong) trion resonance is observed.
   \textbf{(b)} Optical reflectivity spectra from the quasicrystal D$_2$ at several hole doping densities along the black dashed line presented in panel a. 
     }
    \label{fig:fig8}    
\end{figure}

\begin{figure}[h]
 \includegraphics[width=16.5cm]{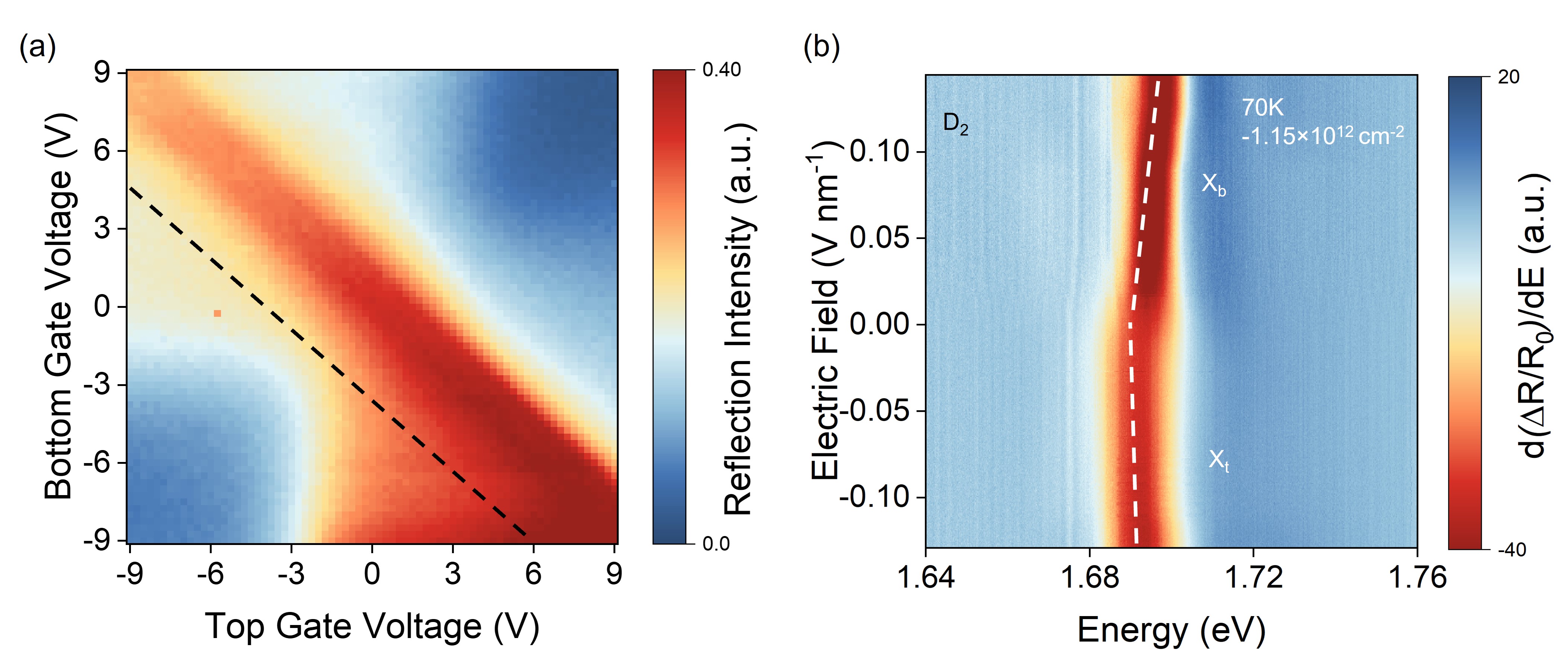}
\caption{\textbf{Reflectivity Spectra at 70K for the hole-doped quasicrystal D$_2$.} 
 \textbf{(a)}
The 1s exciton intensity is plotted as a function of both top and bottom gates at 70K. The E-field dependent reflectivity spectra at a constant hole doping -1.15$\times 10^{12}cm^{-2}$ along the black dashed line are shown at Fig.~S7b.
 \textbf{(b)}
 Reflection spectra depicted as a function of electric field strength with a constant hole doping of approximately -1.15$\times10^{12}$cm$^{-2}$ at 70K. The trion signal becomes much weaker which is consistent with previous work~\cite{arora2019excited}. No abrupt changes with E-field are observed due to thermal broadening.
    }
    \label{fig:fig9}    
\end{figure}

\begin{figure}[h]
 \includegraphics[width=16.5cm]{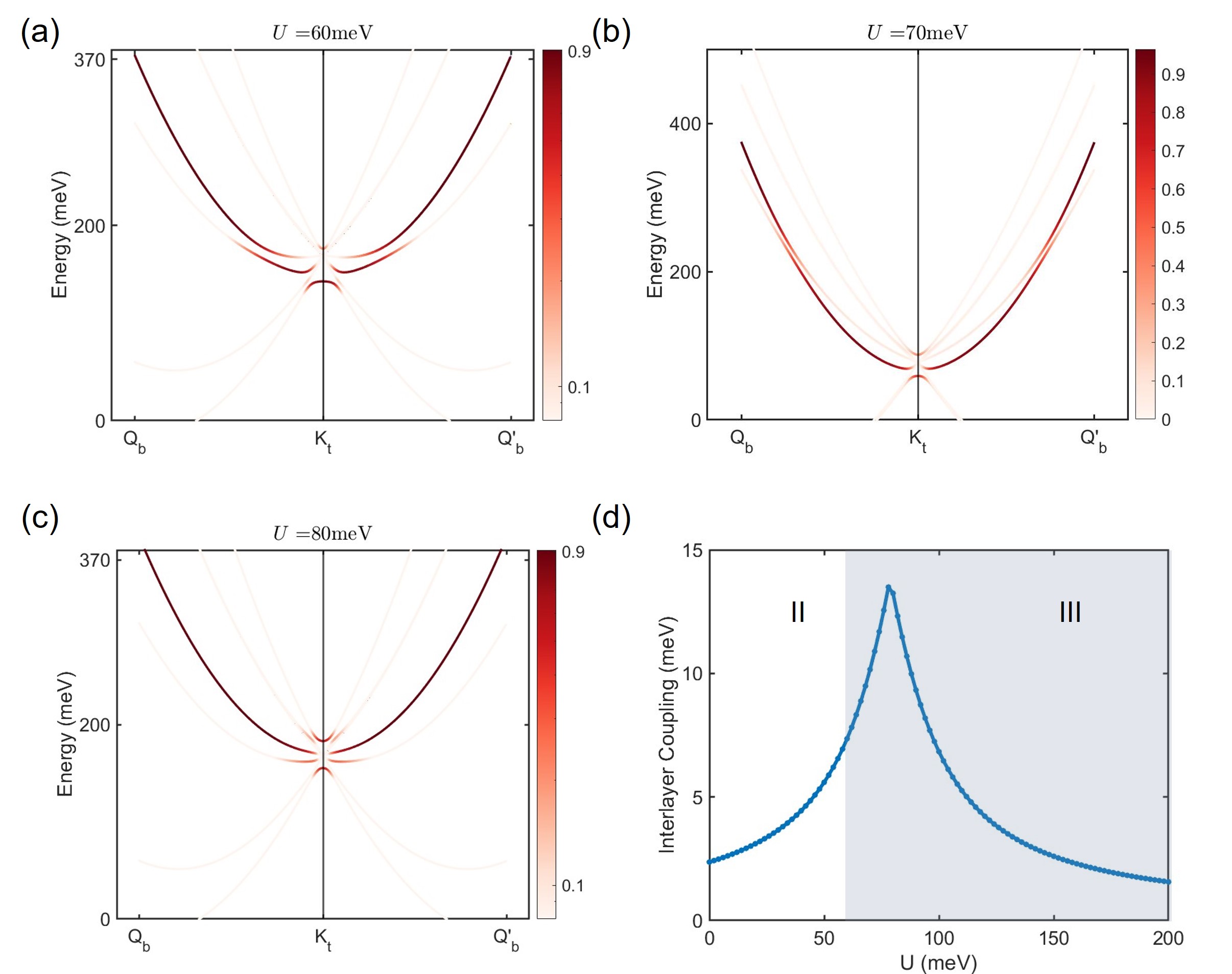}
\caption{\textbf{ Calculated $K_t$ and $Q_b$ hybridization at several  interlayer energy difference $U$ taking into account the third-order Umklapp scattering.} 
\textbf{(a-c)} Calculated hybrid $K-Q$ states at an interlayer energy difference $U$ of 60, 70 and 80 meV. The oscillation strength of lower hybridized state becomes smaller and the higher hybridized state becomes larger as $U$ increases.
\textbf{(d)} The effective interlayer $K-Q$ coupling strength characterized by the energy correction to the $K_t$ valley energy due to the existence of the nearby $Q_b$ valleys and the displacement field.}
    \label{fig:fig11}    
\end{figure}

\end{document}